\documentclass[11pt]{iopart}
\pdfoutput=1
\usepackage{cite,amssymb,amsfonts,graphicx,bm,dcolumn,mathrsfs}
\usepackage{color}
\usepackage{ulem}
\usepackage[utf8]{inputenc}
\usepackage[T1]{fontenc}
\usepackage{lmodern}

\begin{document}

\title{Lifetime of locally stable states near a phase transition in the Thirring model}
\vspace{5mm}
\author{Elaheh Saadat$^1$, Ivan Latella$^2$ and Stefano Ruffo$^{1,3}$}

\address{$^1$ SISSA, via Bonomea 265 and INFN, Sezione di Trieste, 34136 Trieste, Italy}
\address{$^2$ Departament de F\'isica de la Mat\`eria Condensada, Universitat de Barcelona, \\ Mart\'i i Franqu\`es~1, 08028 Barcelona, Spain}
\address{$^3$ Istituto dei Sistemi Complessi, Consiglio Nazionale delle Ricerche, Via Madonna del Piano 10, 50019 Sesto Fiorentino, Italy}

\ead{\mailto{esaadat@sissa.it}, \mailto{ilatella@ub.edu} and \mailto{ruffo@sissa.it}}

\begin{abstract}
We study the lifetime of locally stable states in the Thirring model, which describes a system of particles whose interactions are long-range.
The model exhibits first-order phase transitions in the canonical ensemble and, therefore, a free energy barrier separates two free energy minima. 
The energy of the system diffuses as a result of thermal fluctuations and we show that its dynamics can be described by means of a Fokker-Planck equation. 
Considering an initial state where the energy takes the value corresponding to one of the minima of the free energy, we can define the lifetime of the initial state as the mean first-passage time for the system to reach the top of the free energy barrier between the minima.
We use an analytical formula for the mean first-passage time which is based on the knowledge of the exact free energy of the model, even at a finite number of particles. This formula shows that the lifetime of locally stable states increases exponentially in the number of particles, which is a typical feature of systems with long-range interactions.
We also perform Monte Carlo simulations in the canonical ensemble in order to obtain the probability distribution of the first-passage time, which turns out to be exponential in time in a long time limit. The numerically obtained mean first-passage time agrees with the theoretical prediction.
Combining theory and simulations, our work provides a new insight in the study of metastability in many-body systems with long-range interactions.
\end{abstract}


\maketitle

\section{Introduction}

The study of systems with long-range interactions attracted considerable interest in recent years~\cite{Campa_2014,Campa_2009,Levin_2014,Bouchet_2010,Feliachi_2022}. Their behavior in the framework of statistical mechanics both in equilibrium and out-of-equilibrium has been carefully studied. In these systems, the interaction between two particles slowly decays as $r^{-\alpha}$ at large distances $r$, with $\alpha\leq d$, $d$ being the dimension of the embedding space. This makes the range of the interactions comparable to the size of the system even if the size tends to infinity, introducing interesting features that are absent when the interactions are short-range. Examples of such systems include plasmas~\cite{Kiessling_2003}, two-dimensional flows~\cite{Miller_1990,Robert_1991,Chavanis_2002_b,Bouchet_2009,Venaille_2012}, systems with wave-particle interactions~\cite{Barre_2004,Barre_2005}, self-gravitating
systems~\cite{Lynden-Bell_1968,Thirring_1970,Padmanabhan_1990,Lynden-Bell_1999,Chavanis_2002,Chavanis_2006}, long-range Hamiltonian models~\cite{Antoni_2022,Dauxois_2003} and simple spin models with mean-field interactions~\cite{Barre_2001,Mukamel_2005}. 

A salient property of long-range interacting systems is that they are intrinsically non-additive, leading to the possibility of ensemble inequivalence~\cite{Thirring_1970,Padmanabhan_1990,Ellis_2000,Barre_2001,Bouchet_2005,Chavanis_2006} and to the emergence of an additional term in the Gibbs-Duhem relation~\cite{Latella_2013,Latella_2015,Latella_2017,Campa_2018}. Their dynamics also presents interesting features, since these systems may remain trapped in long-lived quasi-stationary states~\cite{Latora_1998,Yamaguchi_2004,Miller_2023} before evolving towards equilibrium, and exhibit  non-equilibrium phenomena such as anomalous relaxation and diffusion~\cite{Latora_1999,Yamaguchi_2003,Pluchino_2004,Bouchet_2005_b,Yamaguchi_2007}. Interestingly, the relaxation timescale of non-equilibrium quasi-stationary states (which are stable under Vlasov dynamics) increases algebraically with  $N$, the number of particles in the system~\cite{Yamaguchi_2004,Kavita_2007,Chavanis_2012}. 
The scaling of the relaxation time is not algebraic in $N$, however, if the system is found initially in a metastable state corresponding to a local minimum of the free energy, this being the situation that concerns us here.

By considering a Curie-Weiss Ising model in an external magnetic field, it was shown in~\cite{Griffiths_1966} that the relaxation time of a metastable state in the canonical ensemble scales with the exponential of $N$, which was also verified theoretically and with numerical simulations in the microcanonical ensemble for a model that describes the motion of particles with all-to-all interactions in a two-dimensional bounded domain~\cite{Antoni_2004}. The same scaling was obtained for self-gravitating systems in both the microcanonical and canonical ensemble~\cite{Chavanis_2005} by using an adaptation of the Kramers formula as well as in the Keller-Segel model describing bacterial populations that undergo chemotaxis~\cite{Chavanis_2014}.
Since this relaxation time quantifies the lifetime of the metastable state, an important consequence of the exponential scaling is that such states are robust and, therefore, cannot be disregarded~\cite{Chavanis_2005}. 

Here, we analyze the lifetime of locally stable states near a first-order phase transition in the Thirring model~\cite{Thirring_1970}, which is a simplified version of a self-gravitating gas and a workable example to explore the different phenomena observed in systems with long-range interactions~\cite{Campa_2016,Campa_2020,Campa_2022,Trugilho_2022}. 
We recall an analytical formula for the mean first-passage time depending on the exact free energy of the model which is known even for a finite number of particles. This formula makes clear that lifetime of locally stable states, including metastable states, increases exponentially with the number of particles.
By performing Monte Carlo (MC) simulations in the canonical ensemble of the Thirring model close to the first-order phase transition, we compute the first-passage time accounting for the time needed by the system to escape from a local minimum of free energy and reach the top of the barrier separating the two phases. We not only retrieve lifetimes of locally stable states increasing exponentially with $N$ on average, which are identified as the mean first-passage time in our simulations, but we also sample individual first-passage times and find that their distribution is approximately exponential in time in the long time limit. 

In section~\ref{sec:theory} we outline the theoretical approach describing the first-passage time problem in terms of energy diffusion, while in section~\ref{sec:model} we define the model for which this theoretical framework is applied. In section~\ref{sec:simulations} we compare the theory with MC simulations and, finally, in section~\ref{sec:conclusions} we present our conclusions.

\section{Energy diffusion and lifetime of locally stable states}
\label{sec:theory}

Near a first-order phase transition in the canonical ensemble, the free energy exhibits two local minima which correspond to two locally stable states. To quantify the lifetime of these states, in section~\ref{sec:theory_1} we first look for the Fokker-Planck equation describing the evolution of the energy when the system is in contact with a thermal bath. From here, in section~\ref{sec:theory_2} we obtain the mean first-passage time for the system to cross the free energy barrier separating the two locally stable states, assuming that initially the system is in one of these states.

\subsection{Fokker-Planck equation for the energy}
\label{sec:theory_1}

Consider a system with $N$ particles of mass $m$ enclosed in a volume $V$ whose energy $E$ is given by the Hamiltonian
\begin{equation}
\mathcal{H}=\sum_{i=1}^N\frac{\mathbf{p}_i^2}{2m} +W(\mathbf{x}_1,\dots,\mathbf{x}_N) ,
\label{Hamiltonian}
\end{equation}
where $\mathbf{x}_i$ and $\mathbf{p}_i$ are the position and momentum of the $i$-th particle, respectively, $i=1,\dots N$. Here $W(\mathbf{x}_1,\dots,\mathbf{x}_N)$ is the potential energy of the system, which, in particular, will be considered to arise from long-range interactions. Assuming that the interaction potential is properly regularized at short distances and $N$ is finite, there exists a lower bound $E_*$ for the energy of the system such that $E\geq E_*$. 
In addition, the number of microstates with energy $E\geq \mathcal{H}$ is
\begin{eqnarray}
I(E)&=\frac{1}{h^{3N}N!}\int \mathrm{d}^{3N}x\, \mathrm{d}^{3N}p\, \Theta(E-\mathcal{H}),
\label{microstates}
\end{eqnarray}
where $h$ is a constant and $\Theta(x)$ is the step function, whereas the density of states with energy $E$ is obtained as
\begin{equation}
g(E)=\frac{\mathrm{d}I(E)}{\mathrm{d}E}. 
\end{equation}
As we will see, $I(E)$ and $g(E)$ are useful to define properties of the system in situations out of equilibrium.

Now consider the system in a bath with thermal noise at temperature $T$.
The equations of motion of the particles are given by
\begin{eqnarray}
\frac{\mathrm{d}  \mathbf{x}_i}{\mathrm{d}t} =\frac{\mathbf{p}_i}{m},\\
\frac{\mathrm{d}  \mathbf{p}_i}{\mathrm{d}t} = -\frac{\partial}{\partial \mathbf{x}_i} W(\mathbf{x}_1,\dots,\mathbf{x}_N)- \zeta\frac{\mathbf{p}_i}{m} + \mathbf{F}_p(t),
\end{eqnarray}
where $\zeta$ is the friction coefficient and $\mathbf{F}_p(t)$ is the stochastic force satisfying the fluctuation-dissipation theorem
\begin{equation}
\langle \mathbf{F}_p(t)  \mathbf{F}_p(t') \rangle=2\zeta k_BT\delta(t-t'),
\end{equation}
where $k_B$ is the Boltzmann constant.
From the above Langevin equations, the associated $N$-body Fokker-Planck equation for the distribution function $P_N$ is~\cite{Chavanis_2005}
\begin{equation}
\frac{\partial P_N}{\partial t} +\sum_{i=1}^N \left( \frac{\mathbf{p}_i}{m} \frac{\partial P_N}{\partial \mathbf{x}_i}
+ \mathbf{F}_i \frac{\partial P_N}{\partial \mathbf{p}_i}\right)
=\sum_{i=1}^N\frac{\partial}{\partial \mathbf{p}_i}\left[ \zeta k_B T  \frac{\partial P_N}{\partial \mathbf{p}_i} + \zeta P_N \frac{\mathbf{p}_i}{m}\right],
\label{Full_FPE}
\end{equation}
where
\begin{equation}
\mathbf{F}_i=-\frac{\partial}{\partial \mathbf{x}_i} W(\mathbf{x}_1,\dots,\mathbf{x}_N)
\end{equation}
here is the long-range force acting on the $i$-th particle. We now simplify the problem and look for an approximate equation for the probability by following the procedure described in~\cite{Chavanis_2005}. The distribution function is taken as $P_N(\{\mathbf{x}_i,\mathbf{p}_i\})=\bar{P}_N(E,t)$ and is replaced in equation~(\ref{Full_FPE}), obtaining an equation for $\bar{P}_N(E,t)$. This equation is subsequently averaged over the hypersurface of energy $E$ which removes the dependence on the coordinates (microcanonical average). Hence, taking the distribution of energies as $P(E,t)=g(E)\bar{P}_N(E,t)$, one arrives at the Fokker-Planck or Smoluchowski equation~\cite{Chavanis_2005}
\begin{equation}
\frac{\partial}{\partial t}  P(E,t)= \frac{\partial}{\partial E}  D(E)\left( \frac{\partial }{\partial E}P(E,t)+\beta P(E,t) \frac{\partial }{\partial E}F(E)\right),
\label{eq_P}
\end{equation}
where $\beta=1/k_BT$, $F(E)=E -k_B T\ln g(E)$ is the free energy\footnote{The function $F(E)$ is actually the free energy of a system in equilibrium when evaluated at a local minimum (the most probable state). The energy at the minimum is a function of the temperature, so at this point, in fact, we have $F=F(T)$.} for a given energy $E$, and the energy diffusion coefficient is given by
\begin{equation}
D(E)= 3N k_BT \frac{\zeta}{m}\frac{I(E)}{g(E)}.
\label{energy_diff_coef}
\end{equation}
The equilibrium probability corresponding to the stationary solution of equation~(\ref{eq_P}) is
\begin{equation}
P_\mathrm{eq}(E)=\frac{1}{Z(\beta)}g(E)e^{-\beta E} =\frac{1}{Z(\beta)}e^{-\beta F(E)},
\end{equation}
where 
\begin{equation}
Z(\beta)= \frac{1}{\Lambda^{3N}N!}\int d^{3N}x\, e^{-\beta W(\mathbf{x}_1,\dots,\mathbf{x}_N)}
\label{partition_fucntion}
\end{equation}
is canonical partition function, $\Lambda=h\sqrt{\beta/2\pi m }$ being the thermal wavelength.
Equation~(\ref{eq_P}) describes how energy diffuses in time when the system is brought out of equilibrium. 

\subsection{Mean first-passage time
\label{sec:theory_2}}

Here we are interested in the characterization of the lifetime of locally stable states in a system undergoing a first-order phase transition. The evolution of the system through the free energy barrier separating the two phases can be discerned by using the energy as the parameter defining the state of the system, and the energy dynamics is precisely dictated by the Fokker-Planck equation (\ref{eq_P}). In the situation of interest, we assume that initially the system has energy $E_A$ at a local minimum of free energy and that weak noise and friction perturb the state of the system. Even for weak noise, the energy of the system is affected by fluctuations and the system eventually escapes from the free energy well. For the phase transition to occur, the system must cross a free energy barrier whose maximum is located at an energy $E_B$. The time it takes for the system with initial energy $E_A$ to reach the top of the barrier for the first time corresponds to a certain time $t$. Hence, the lifetime of the locally stable state $t_\mathrm{life}$ can be quantified, on average, by the mean first-passage time $\tau(E_A)=\langle t\rangle$, that is
\begin{equation}
t_\mathrm{life}= \tau(E_A).
\label{t_life}
\end{equation}

The mean first-passage time for the considered situation can be deduced from the Fokker-Planck equation (\ref{eq_P}) by taking into account an initial condition for the probability and suitable boundary conditions~\cite{Zwanzig,Szabo_1980,Hanggi_1990,Berezhkovskii_2019}. Let $P(E,t|E_0)$ be the probability density of finding the system with energy $E$ at time $t$, given that at $t=0$ the energy of the system was $E_0$ (not necessarily at a local minimum of the free energy). Equation (\ref{eq_P}) for $P(E,t|E_0)$ can be recast into
\begin{equation}
\frac{\partial}{\partial t}  P(E,t|E_0)= \mathcal{L}_E P(E,t|E_0)
\label{Fokker-Planck}
\end{equation}
and solved with the initial condition $P(E,0|E_0)=\delta(E-E_0)$, where we have introduced the Fokker-Planck operator
\begin{equation}
\mathcal{L}_E= -\frac{\partial}{\partial E} v(E)
+  \frac{\partial}{\partial E}  D(E)\frac{\partial }{\partial E}.
\end{equation}
Here the drift is given by
\begin{equation}
v(E)=-D(E)\beta\frac{\partial}{\partial E}F(E)=D(E)\left(\beta_E -\beta\right),
\end{equation}
where
\begin{equation}
\beta_E=\frac{\partial}{\partial E}\ln g(E) 
\end{equation}
is the inverse microcanonical temperature. We observe that the drift is due to the difference of the temperature of the bath and that corresponding to the isolated system in equilibrium at the given energy $E$. 

Let us now denote by $\Omega$ and $\partial\Omega$ the domain of energies for the Fokker-Planck equation and its boundary, respectively, which will be defined below for the problem at hand. The survival probability $S(E_0,t)$ for the system at time $t$ with initial energy $E_0$ is obtained by integrating the probability density over this domain,
\begin{equation}
S(E_0,t)=\int_\Omega dE\, P(E,t|E_0).
\end{equation}
The distribution of first-passage times is then given by~\cite{Zwanzig}
\begin{equation}
\rho(E_0,t)=-\frac{\partial}{\partial t} S(E_0,t),
\end{equation}
in such a way that the mean first-passage time is the first moment of $t$,
\begin{equation}
\tau(E_0)=\int_0^\infty dt\, t \rho(E_0,t) .
\end{equation}
Moreover, integrating the above equation by parts and using that $S(E_0,t)$ vanishes for $t\to\infty$ yields
\begin{equation}
\tau(E_0)=\int_0^\infty dt\, S(E_0,t) .
\label{mfpt_S}
\end{equation}

A useful equation to determine $\tau(E_0)$ can be found by considering the operator that is adjoint to $\mathcal{L}_E$ acting on the initial energy $E_0$, given by
\begin{equation}
\mathcal{L}^\dagger_{E_0}= v(E_0)\frac{\partial}{\partial E_0} 
+  \frac{\partial}{\partial E_0}  D(E_0)\frac{\partial }{\partial E_0}.
\label{adjoint_operator}
\end{equation}
Thus, it can be shown that operating with $\mathcal{L}^\dagger_{E_0}$ on equation~(\ref{mfpt_S}) yields~\cite{Zwanzig}
\begin{equation}
\mathcal{L}^\dagger_{E_0}  \tau(E_0)=-1,
\label{condition_MFPT_operator}
\end{equation}
which must be solved taking into account the appropriate conditions on $\partial\Omega$. An absorbing boundary condition implies that $\tau(E_0)$ vanishes for $E_0$ on $\partial\Omega$, while the derivative $\partial_{E_0} \tau(E_0)$ vanishes for $E_0$ on $\partial\Omega$ when a reflecting boundary condition is imposed~\cite{Hanggi_1990,Gardiner}.

We now suppose that the system is initially in the low-energy phase, so that $\Omega$ is defined by the lower bound $E_*$, the minimum energy that the system can achieve, and the energy $E_B$ where  the free energy barrier is located. To compute $\tau(E_0)$, a reflecting boundary condition is taken on $E_*$ and an adsorbing boundary condition is imposed on $E_B$.
Since the adjoint operator (\ref{adjoint_operator}) can also be written as
\begin{equation}
\mathcal{L}_{E_0}^\dagger= e^{\beta F(E_0)}\frac{\partial}{\partial E_0}
  e^{-\beta F(E_0)}D(E_0)\frac{\partial }{\partial E_0} ,
\end{equation}
operating according to (\ref{condition_MFPT_operator}) and taking into account the boundary conditions leads to
\begin{equation}
\tau(E_0)= \int^{E_B}_{E_0}dE\frac{e^{\beta F(E)}}{D(E)} \int_{E_*}^{E}dE' e^{-\beta F(E')}.
\label{MFPT_low_energy}
\end{equation}
Analogously, for the case in which the system is in the high-energy phase, the domain of energies is $[E_B,\infty]$ and a reflecting boundary condition is taken at infinity. In this case one finds
\begin{equation}
\tau(E_0)= \int_{E_B}^{E_0}dE\frac{e^{\beta F(E)}}{D(E)} \int^{\infty}_{E}dE' e^{-\beta F(E')}.
\label{MFPT_high_energy}
\end{equation}
Thus, if the initial energy is that of a free energy minimum, $ E_0=E_A$, the lifetime of a locally stable state $t_\mathrm{life}= \tau(E_A)$ is obtained using (\ref{MFPT_low_energy}) for the low-energy phase ($E_A<E_B$) and using (\ref{MFPT_high_energy}) for the high-energy phase  ($E_B<E_A$). 

A similar result for $t_\mathrm{life}$ was obtained in~\cite{Chavanis_2005} for self-gravitating systems by using an adaptation of the Kramers formula and explicitly obtaining an approximate expression the probability $P(E,t)$. In contrast, the method above does not require finding the probability. Moreover,
by expanding the free energy in powers of $E$ and approximating the integrals in (\ref{MFPT_low_energy}) or (\ref{MFPT_high_energy}) by a Gaussian, it is found that~\cite{Chavanis_2005}
\begin{equation}
t_\mathrm{life} \simeq \frac{\pi \sqrt{C(E_A)|C(E_B)|}}{\beta^2 D(E_B)}e^{\beta [F(E_B)-F(E_A)]},
\end{equation}
where $C(E_A)>0$ and $C(E_B)<0$ are the heat capacities of the system in the locally stable and unstable states, respectively. Since the height of the free energy barrier $\beta [F(E_B)-F(E_A)]$ scales as $N$, the lifetime of a locally stable state scales as $e^N$. As mentioned in the introduction, the same scaling is obtained in~\cite{Griffiths_1966,Antoni_2022} for metastable states from considerations on the magnetization in spin systems.

To apply the method explained above, in the following section we introduce the Thirring model for which we obtain the lifetime of locally stable states as a function of the number of particles both theoretically and with MC simulations. 

\section{Thirring model}
\label{sec:model}

We consider a model introduced by Thirring~\cite{Thirring_1970} to describe a simplified version of a self-gravitating system. The model exhibits several interesting properties such as ensemble inequivalence and, in particular, a first-order phase transition in the canonical ensemble~\cite{Campa_2016}. It consists of $N$ particles of mass $m$ enclosed in a volume $V$ which has an internal region of volume $V_c$ defining a ``core''. Each particle in the core interacts equally with all other particles in the core with a constant, attractive potential, while particles outside this region are free.  If for a given configuration the number of particles in the core is $N_c$, the total potential energy of the system is
\begin{equation}
W(N_c)=-\nu N_c(N_c-1) ,
\label{potential_energy}
\end{equation}
where $\nu>0$ is a coupling constant. When the number of particles is large, the potential energy behaves as $-\nu N_c^2$. Here the Hamiltonian $\mathcal{H}$ is given by (\ref{Hamiltonian}) with the potential energy (\ref{potential_energy}). We now derive for this model all quantities specified in the previous section needed to compute the lifetime of locally stable states near the phase transition.

The number of microstates with energy $\mathcal{H}\leq E$, defined in (\ref{microstates}), for this model can be written as~\cite{Thirring_1970}
\begin{equation}
I(E)=A\sum^{N}_{N_c=N_\mathrm{min}}(E-W)^{3N/2}\frac{e^{\eta (N-N_c)}}{N_c!(N-N_c)! },
\end{equation}
where $A=V_c^N(2m\pi/h^2)^{3N/2}/(3N/2)!$ and $\eta=\ln(V/V_c-1)$ represents a reduced volume.
The number of particles $N_\mathrm{min}$ corresponds to the minimum value of $N_c$ such that $E-W>0$, ensuring that the kinetic energy is non-negative.
In addition, we restrict the domain of energies to $E\geq E_*$ with the lower bound $E_*=-\nu N(N-1)$ corresponding to all particles in the core with a vanishing kinetic energy. Taking this bound into account,
we introduce the reduced energy
\begin{equation}
\varepsilon=\frac{E-E_*}{\nu N^2} \geq0,
\end{equation}
which remains finite in the large $N$ limit because the potential energy scales as $N^2$.
The number of microstates can then be written as
\begin{equation}
I(E)=(\nu N^2)^{3N/2}A\tilde{I}(\varepsilon) ,
\end{equation}
where we have introduced
\begin{equation}
\tilde{I}(\varepsilon)=\sum^{N}_{N_c=N_\mathrm{min}}(\varepsilon - w)^{3N/2}\frac{e^{\eta (N-N_c)}}{N_c!(N-N_c)! }
\end{equation}
with $w=  ( N+N_c - 1 )(N-N_c)/ N^2$. Furthermore, the density of states is given by $g(E)=\mathrm{d}I(E)/\mathrm{d}E$ and reads
\begin{equation}
g(E)=\frac{3}{2}N(\nu N^2)^{3N/2-1}A \tilde{g}(\varepsilon),
\end{equation}
where
\begin{equation}
\tilde{g}(\varepsilon)=\sum^{N}_{N_c=N_\mathrm{min}}(\varepsilon - w)^{3N/2-1}\frac{e^{\eta (N-N_c)}}{N_c!(N-N_c)! }.
\end{equation}
We also introduce a reduced free energy $f(\varepsilon)$ according to $F(E)=\nu N^2f(\varepsilon)+F_*$, where 
\begin{equation}
f(\varepsilon)=\varepsilon-\frac{\theta}{N}\ln\tilde{g}(\varepsilon),
\label{free_energy}
\end{equation}
and we have separated a contribution $F_*=E_*-k_BT\ln[3N(\nu N^2)^{3N/2-1}A/2]$ which does not depend on $\varepsilon$. In equation (\ref{free_energy}) we have defined the reduced temperature
\begin{equation}
\theta=\frac{k_B T}{\nu N},
\end{equation}
which remains finite in the large $N$ limit following the usual scaling in long-range interacting systems. 

\begin{figure}
\centering 
\includegraphics[scale=1]{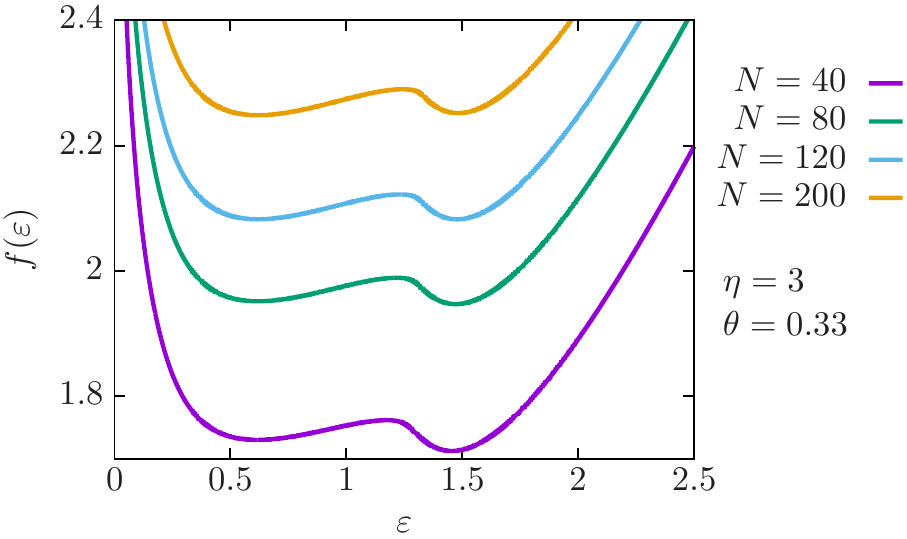}
\caption{Reduced free energy as a function of the reduced energy for different number of particles, taking the reduced volume and temperature as $\eta=3$ and $\theta=0.33$, respectively.}
\label{fig_free_energy}
\end{figure}

The model exhibits first-order phase transitions in the canonical ensemble that terminate at a critical point defined by $\eta_\mathrm{cp}=2$ and $\theta_\mathrm{cp}=1/2$, and these transitions occur for $\eta>\eta_\mathrm{cp}$ and $\theta<\theta_\mathrm{cp}$~\cite{Campa_2016}. 
In figure~\ref{fig_free_energy}, we show the reduced free energy as a function of the reduced energy for different number of particles in a typical configuration studied here. For a fixed reduced volume, we chose the reduced temperature such that the system is near the phase transition, so in this example $\eta=3$ and $\theta=0.33$. In figure~\ref{fig_free_energy}, the local minimum at low energies corresponds to a ``condensed'' phase characterized by a large number of particles in the core, while the minimum at high energies corresponds to a ``dilute'' phase. 
Below we also illustrate the dependence of the phase transitions with respect to the temperature, see figure~\ref{fig_n0_vs_T}.

Furthermore, according to (\ref{energy_diff_coef}), the energy diffusion coefficient for this model takes the form
\begin{equation}
D(E) = k_B T\frac{2\zeta}{m} \nu N^2 q(\varepsilon),
\end{equation}
where $q(\varepsilon)=\tilde{I}(\varepsilon)/\tilde{g}(\varepsilon)$.
Hence, the lifetime $t_\mathrm{life}= \tau(\varepsilon_A)$ given by equation (\ref{MFPT_low_energy}) for the condensed phase can be written as
\begin{equation}
\frac{t_\mathrm{life}}{t_c}= N\int_{\varepsilon_A}^{\varepsilon_B}\mathrm{d}\varepsilon\,  \frac{e^{Nf(\varepsilon)/\theta}}{\theta q(\varepsilon)} \int^{\varepsilon}_0 \mathrm{d}\varepsilon' e^{- Nf(\varepsilon')/\theta}  ,
\label{mfpt_condensed}
\end{equation}
where $\varepsilon_A$ and $\varepsilon_B$ are the reduced energies corresponding to the minimum and maximum of the free energy, respectively, and we have introduced the characteristic time
\begin{equation}
t_c=\frac{ m}{2 \zeta }. 
\end{equation}
Similarly, using (\ref{MFPT_high_energy}) for the dilute phase one has
\begin{equation}
\frac{t_\mathrm{life}}{t_c}= N\int^{\varepsilon_A}_{\varepsilon_B}\mathrm{d}\varepsilon\,  \frac{e^{Nf(\varepsilon)/\theta}}{\theta q(\varepsilon)} \int_{\varepsilon}^\infty \mathrm{d}\varepsilon' e^{- Nf(\varepsilon')/\theta}  .
\label{mfpt_dilute}
\end{equation}
Note that $\varepsilon$, $\theta$, $f$, and $q$ in expressions (\ref{mfpt_condensed}) and (\ref{mfpt_dilute}) are dimensionless and of order unity in $N$. Although in the simulations below we describe finite systems, we take the number of particles large enough such that the scaling corresponding to the large $N$ limit is meaningful. 

\section{Simulations}
\label{sec:simulations}

\begin{figure}
\centering
\includegraphics[scale=1]{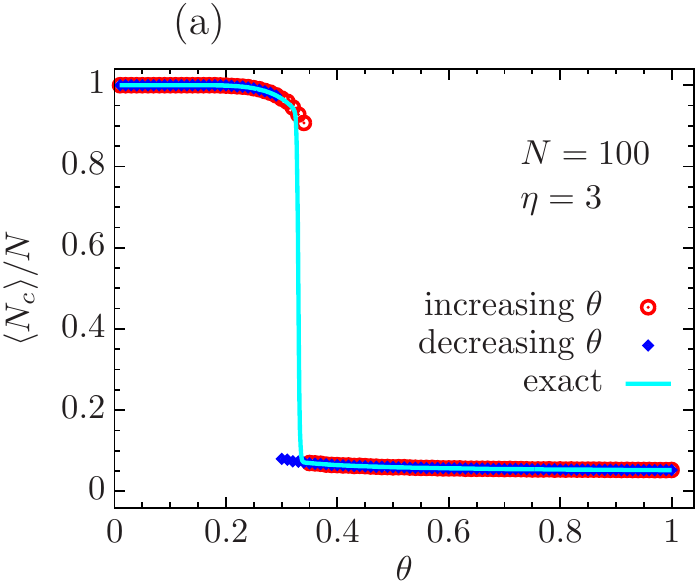}~\hspace{2mm}~\includegraphics[scale=1]{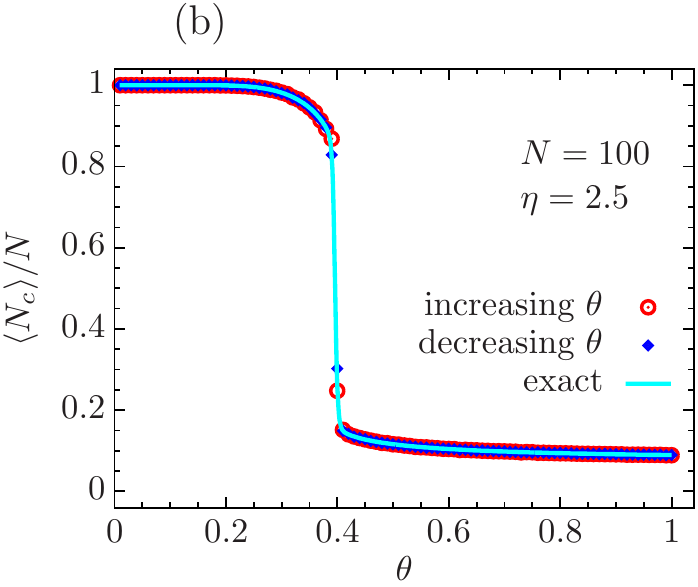}
\caption{Averaged fraction of number of particles in the core as a function of the reduced temperature for $\eta=3$ in (a) and $\eta=2.5$ in (b). Symbols represents the results of MC simulations and solid lines the exact result obtained from equation~(\ref{fraction}).}
\label{fig_n0_vs_T}
\end{figure}

As explained in the preceding section, the system can be in either a condensed or a dilute phase. To determine the actual phase in a given configuration, one can measure the fraction of particles in the core which acts as an order parameter~\cite{Campa_2016,Campa_2020}. For finite number of particles, this fraction can be obtained exactly via numerical computation. One way to proceed involves the canonical partition function (\ref{partition_fucntion}) which for the Thirring model can be written as~\cite{Latella_2017}
\begin{equation}
Z(N,\eta,\theta)= \left(\frac{V_c}{\Lambda^3}\right)^N \sum_{N_c=0}^N
\frac{e^{\eta (N-N_c) - \mathcal{U}(N_c)/\theta}}{N_c!(N-N_c)!} ,
\end{equation}
where $\mathcal{U}(N_c)=W(N_c)/N$ is the potential energy per particle. 
In view of the above expression, we identify the probability $p(N_c)$ of finding the system with $N_c$ particles in the core for given $N$, $\eta$ and $\theta$, 
\begin{equation}
p(N_c)=
\frac{e^{\eta (N-N_c) - \mathcal{U}(N_c)/\theta}}{\tilde{Z} N_c!(N-N_c)!},
\label{probability}
\end{equation}
where the normalization constant is given by
\begin{equation}
\tilde{Z}= \sum_{N_c=0}^N
\frac{e^{\eta (N-N_c) - \mathcal{U}(N_c)/\theta}}{N_c!(N-N_c)!}.
\end{equation}
In terms of this probability, the average fraction of particles in the core in equilibrium configurations reads
\begin{equation}
\frac{\langle N_c\rangle}{N} =\sum_{N_c=0}^N \frac{N_c}{N} p(N_c).
\label{fraction}
\end{equation}

\begin{figure}
\centering 
\includegraphics[scale=1]{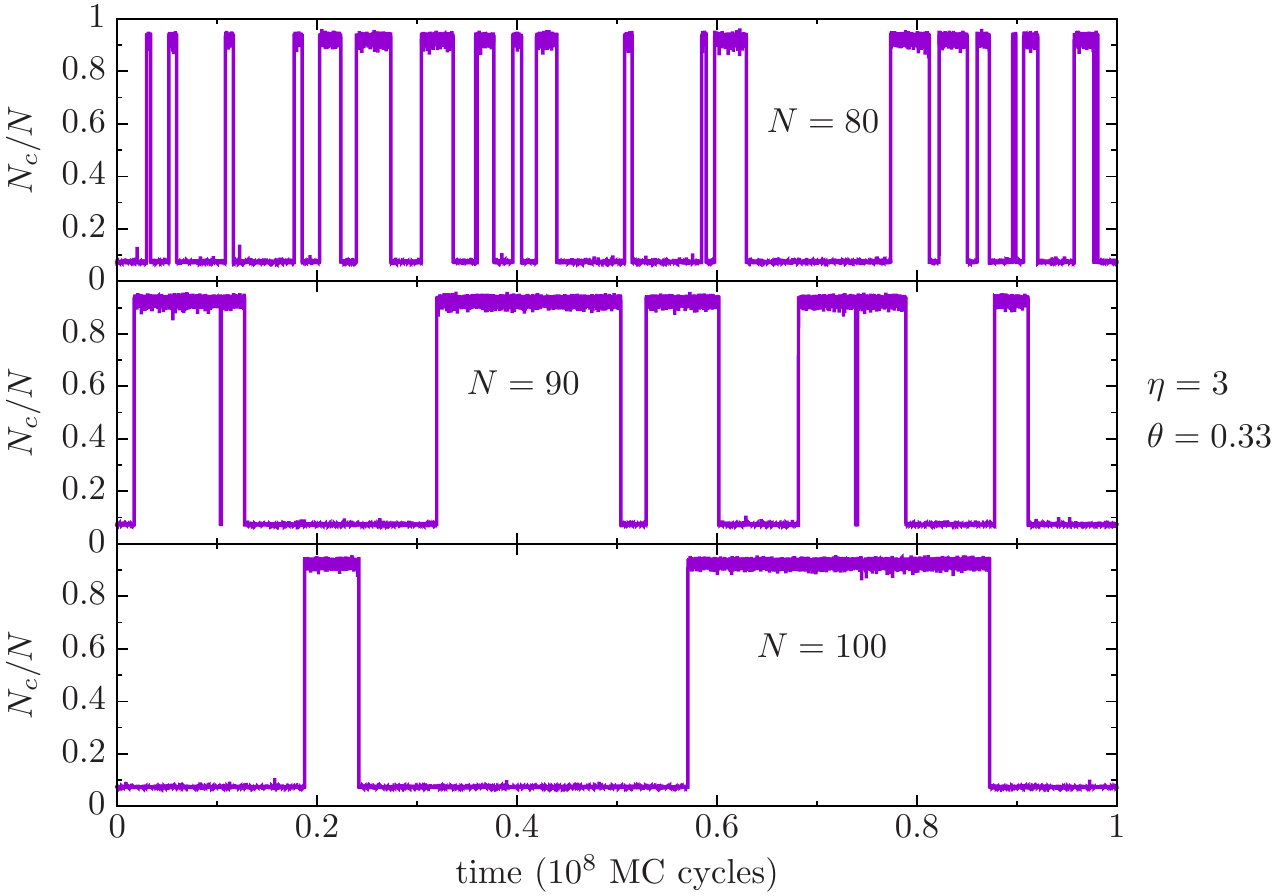}
\caption{Transition between the dilute and condensed phases. We show the evolution of the fraction $N_c/N$ (averaged over a small time window of $10^3$ MC cycles) as a function of time for different number of particles. Here the reduced volume and temperature are $\eta=3$ and $\theta=0.33$, respectively.}
\label{fig_jumps}
\end{figure}

The fraction of particles in the core grows markedly in the condensed phase. In figure~\ref{fig_n0_vs_T}, we plot $\langle N_c\rangle/N$ as a function of the temperature for $N=100$. An abrupt change in this fraction is observed at $\theta$ close to $0.33$ when $\eta=3$ and at $\theta$ close to $0.39$ for $\eta=2.5$.
Obviously $\langle N_c\rangle/N$ is not really discontinuous at these points since the number of particles is finite, but the jump is clearly sharp and can be considered as a phase transition. In the same figure we also show $\langle N_c\rangle/N$ obtained with MC simulations in the canonical ensemble, that we perform throughout this work using the standard Metropolis algorithm~\cite{Frenkel}. As can be seen in the figure, the simulations account for the transitions in accordance with the theory. In addition, if the system is prepared near a phase transition and the simulation is run without changing the parameters, the system experiences repeated transitions between the two phases indefinitely. 
Since the system is finite, in particular, we highlight that transitions from the locally stable state with lower free energy to the metastable state with higher free energy can be realized as well~\cite{Reguera}. 
The time it spends in each phase, that is, the lifetime of the state, increases on average as the number of particles increases. 
This behavior is exemplified in figure~\ref{fig_jumps}. Note that the jumps do not have a constant frequency; we will show below with simulations that the underlying distribution describing the lifetime is approximately exponential. As in the present case, repeated switching between two locally stable states and an exponential distribution of lifetimes were also observed in the stochastic Keller-Segel model describing chemotaxis~\cite{Chavanis_2014}.

We now focus on the characterization of the lifetime of these states.
To perform the simulations we numerically determine the energies $\varepsilon_A$ and $\varepsilon_B$ at which the free energy (\ref{free_energy}) has a local minimum and maximum, respectively, for given $\theta$ and $\eta$. 
Since we chose configurations near the phase transition, there are two different values of $\varepsilon_A$ corresponding to the condensed and dilute phases.
We thus prepare a state with energy $\varepsilon_A$, either in the condensed or dilute phase, and measure the number of MC cycles for which the system reaches the energy $\varepsilon_B$ at the top of the barrier. A MC cycle consists of $N$ elementary moves and here it defines the unit of time. The time needed by the system to reach the energy $\varepsilon_B$ starting at $\varepsilon_A$ is the associated first-passage time $t$. The lifetime of the state is the average of the first-passage time over several realizations, $t_\mathrm{life}= \tau(\varepsilon_A)$, i.e. the first moment of $t$. In our simulations we take $10^3$ realizations. For each realization we perform a calibration stage to set the maximum displacement allowed to the particles, so that the acceptance ratio of MC configurations is fixed at about $50\%$. We emphasize that the different measurement runs are performed always starting from a configuration with energy $\varepsilon_A$.
Furthermore, initial configurations are defined by the number of particles $N_c=N_c^A$ leading to the local minimum of free energy. Since the energy $\varepsilon_A$ is numerically determined by minimization of the free energy, this number of particles follows from the relation
\begin{equation}
E_A=\nu N^2\varepsilon_A -\nu N(N-1)=\frac{3}{2}Nk_BT-\nu N_c^A(N_c^A-1) .
\end{equation}
To prepare the initial configurations, we put $N_c^A$ particles inside the core with a uniform random distribution and $N-N_c^A$ outside the core randomly distributed as well. We recall that since interactions are spatially constant in the core, the density of particles is uniform there for an equilibrium configuration. The densities inside and outside the core are different in general. 

\begin{figure}[!t]
\centering
\includegraphics[scale=1]{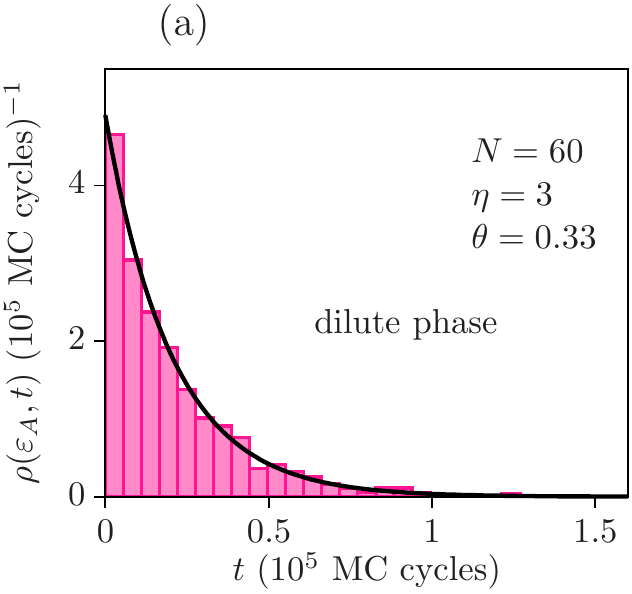}~\hspace{2mm}~\includegraphics[scale=1]{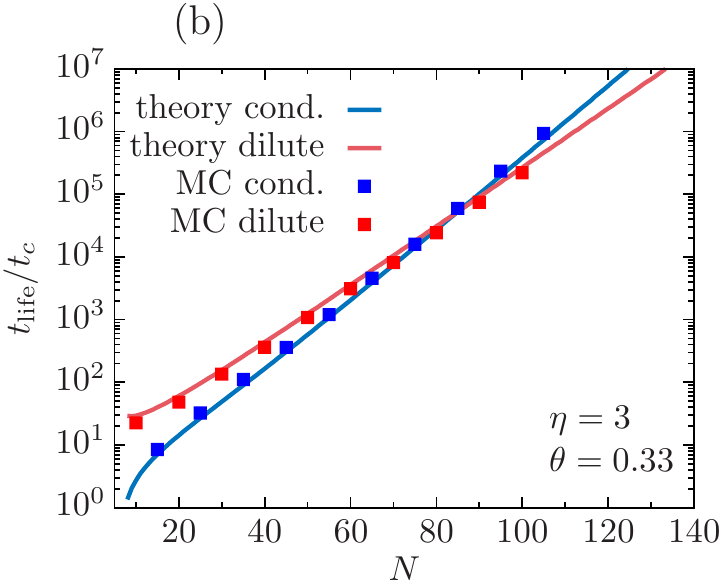}

\vspace{4mm}
\includegraphics[scale=1]{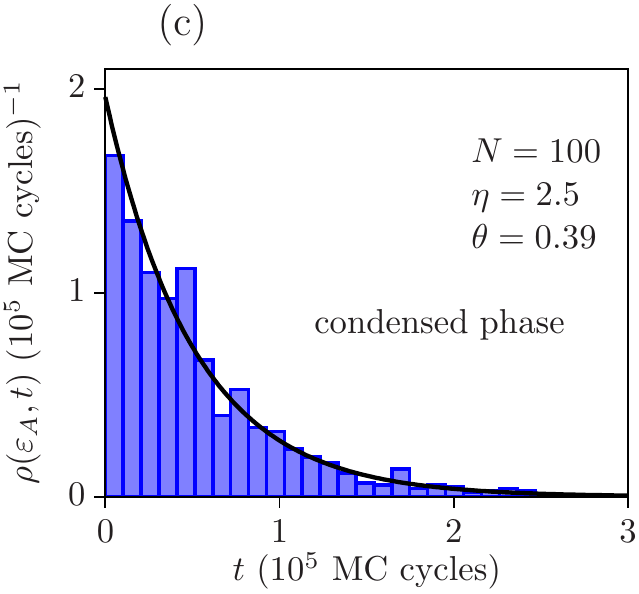}~\hspace{2mm}~\includegraphics[scale=1]{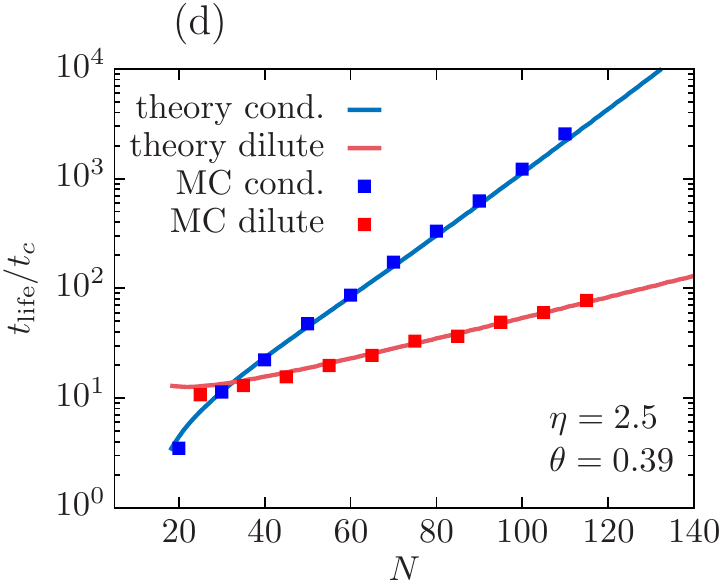}
\caption{Distribution of first-passage times obtained from $10^3$ realization (left panel) and lifetimes as a function of the number of particles  (right panel). In (a) and (b), we set $\eta=3$ and $\theta=0.33$ and fitting the time constant with simulations yields $t_c=(65\pm12)\,$MC cycles. In (c) and (d) we take $\eta=0.25$ and $\theta=0.39$, while the obtained value of the time constant is $t_c=(42\pm4)\,$MC cycles.}
\label{fig_lifetime}
\end{figure}

In order to compare the lifetime measured in the simulations with the theory, we need to determine the characteristic time constant $t_c$ appearing on the left-hand side of equations (\ref{mfpt_condensed}) and (\ref{mfpt_dilute}). To this end, let us denote by $\mathcal{N}_\mathrm{cycles}$ the average number of MC cycles measured in the first-passage time problem explained above, for a given $N$ with $\theta$ and $\eta$ fixed, either in the condensed or dilute phase. Imposing $\mathcal{N}_\mathrm{cycles}=t_\mathrm{life}$ yields
\begin{equation}
t_c=\frac{\mathcal{N}_\mathrm{cycles}}{t_\mathrm{life}/t_c}, 
\end{equation}
where $t_\mathrm{life}/t_c$ in the denominator is given by the right-hand side of equations (\ref{mfpt_condensed}) and  (\ref{mfpt_dilute}) for the condensed and dilute phases, respectively, which can be computed explicitly by numerical integration. For $\theta$ and $\eta$ fixed, we actually take $t_c$ as the average over the results for different values of $N$ for both the condensed and dilute phases. We highlight that in the simulations we do not fix an intrinsic time scale but the acceptance ratio of the MC moves. Thus, with this procedure, it turns out that $t_c$ actually depends on $\theta$ and $\eta$.

In figure~\ref{fig_lifetime}, we show the results of the simulations for two different sets of $\theta$ and $\eta$, chosen in a way that the system is close to the phase transition. The obtained results indicate that the lifetime and its standard deviation coincide for a given $N$ in both the condensed and dilute phases. In agreement with this fact and according to the histograms in figures~\ref{fig_lifetime}(a) and~\ref{fig_lifetime}(c) representing two particular configurations, the obtained first-passage times approximately follow exponential distributions~\cite{Talkner_1987,Sabhapandit_2020}. Solid lines in figures~\ref{fig_lifetime}(a) and~\ref{fig_lifetime}(c) describe the fitting of the first-passage time distribution $\rho(\varepsilon_A,t)=\lambda e^{-\lambda t}$ with the parameter $\lambda=1/\tau(\varepsilon_A)$ obtained from the simulations. As discussed in section~\ref{sec:theory}, this distribution and the survival probability are related through
\begin{equation}
\rho(\varepsilon_A,t)=-\frac{\partial}{\partial t} S(\varepsilon_A,t).
\end{equation}
Thus, simulations indicate that the survival probability is suitably described with an exponential function of the form
\begin{equation}
S(\varepsilon_A,t)=e^{-t/\tau(\varepsilon_A)},
\end{equation} 
as can be expected in a long time limit from a well-behaved distribution~\cite{Masoliver}.

We highlight that once $t_c$ is fitted as discussed above, theory and simulations are in very good agreement, as can be appreciated for $t_\mathrm{life}$ in figures~\ref{fig_lifetime}(b) and~\ref{fig_lifetime}(d). We see that $t_\mathrm{life}/t_c$ in figure~\ref{fig_lifetime}(d) is orders of magnitude smaller than in figure~\ref{fig_lifetime}(b) for the same $N$; the configuration with $\eta=2.5$ and $\theta=0.39$ [figure~\ref{fig_lifetime}(d)] is closer to the critical point than that with $\eta=3$ and $\theta=0.33$ [figure~\ref{fig_lifetime}(b)], so it is easier for the system to change from one phase to the other. Finally, the scaling of $t_\mathrm{life}$ as $e^N$ for this model can be observed in the figures starting at not so large values of $N$.

\section{Summary and conclusions}
\label{sec:conclusions}

We have studied the lifetime of locally stable states, including metastable states, in a long-range interacting system described by the Thirring model~\cite{Thirring_1970}.
We have found that the lifetime of these states increases exponentially with the number of particles, demonstrating, in particular, that metastable states are long-lived.

In the situation we have analyzed, the system is in contact with a thermal bath and a barrier arising from interactions between the particles separates two free energy minima characterizing two locally stable states. Due to thermal fluctuations, the energy of the system diffuses and its dynamics can be described by means of a Fokker-Planck equation. Assuming that initially the system is at a local minimum of the free energy, we have calculated the lifetime of these states as the mean first-passage time taken by the system to reach the top of the barrier. We have performed Monte Carlo simulations in the canonical ensemble to sample the distribution of first-passage times and the lifetime as a function of the number of particles in the system, finding very good agreement with theoretical predictions.

Our results are in accordance with the behavior obtained in other instances of long-range interacting systems~\cite{Griffiths_1966,Antoni_2004,Chavanis_2005,Chavanis_2014}, suggesting that a lifetime increasing exponentially with the number of particles is a rather general feature of metastable states in these systems. This work contributes to the understanding of metastability in systems with long-range interactions.

\section*{Acknowledgments}
We thank R. Klages, S. Majumdar, J. Masoliver and D. Reguera for useful discussions. This work is part of the MIUR-PRIN2017 project Coarse-grained description for nonequilibrium systems and transport phenomena (CO-NEST) No. 201798CZL. I.~L.~acknowledges financial support from the Spanish Government through Grant No. PID2021-126570NB-I00 (MICINN/FEDER, UE).

\end{document}